\documentclass[a4paper,11pt]{article}
\pdfoutput=1 

\usepackage{jcappub} 

\usepackage[T1]{fontenc} 

\usepackage{array}

\newcommand{\hu}{\mathrm{\,km\, Mpc^{-1} s^{-1}}}
\newcommand{\LCDM}{$\Lambda\mathrm{CDM}$ }

\usepackage{multirow}
\usepackage{array}
\newcolumntype{P}[1]{>{\centering\arraybackslash}p{#1}}
\newcolumntype{M}[1]{>{\centering\arraybackslash}m{#1}}

\usepackage{amsmath}

\usepackage{graphicx}
\usepackage{siunitx}
\usepackage{comment}
\usepackage[dvipsnames]{xcolor}

\title{\boldmath Gravitational wave friction in light of GW170817 and GW190521}

\author[a]{S.~Mastrogiovanni,}
\author[a]{L.~Haegel,}
\author[b]{C.~Karathanasis,}
\author[c]{I.~Maga\~{n}a Hernandez,}
\author[a]{D.~A.~Steer}

\affiliation[a]{AstroParticule et Cosmologie (APC), 10 Rue Alice Domon et L\'eonie Duquet, Universit\'e de Paris, CNRS, AstroParticule et Cosmologie, F-75013 Paris, France}

\affiliation[b]{Institut de F´ısica d’Altes Energies (IFAE), Barcelona Institute of Science and Technology, Barcelona, Spain}

\affiliation[c]{Department of Physics, University of Wisconsin-Milwaukee, Milwaukee, WI 53201, USA}

\abstract{We use the gravitational wave (GW) events GW170817 and GW190521, together with their proposed electromagnetic counterparts, to constrain cosmological parameters and theories of gravity beyond General Relativity (GR).
In particular we consider models with a time-varying Planck mass, large extra-dimensions, and  a phenomenological parametrization covering several beyond-GR theories. 
In all three cases, this introduces a friction term into the GW propagation equation, effectively modifying the GW luminosity distance.
We set constraints on \LCDM parameters and GR deviation parameters using two sets of priors on the Hubble constant and matter energy density.
With priors set to the measured Planck's mission values, we find that with the inclusion of GW190521, the two GR deviation parameters constraints improve by a factor $\sim 10$. We report a number of space-time dimensions compatible with $4$ with an precision of $2.5\%$ (at 95\% CL) and an upper limit to the variation of  Newton's constant at the epoch of GW170817 of $<20\%$.  
With wide priors on the Hubble constant and matter energy density, we show that it is still possible to constrain the \LCDM parameters and GR deviation parameters conjointly from GW170817 and GW190521, obtaining constraints on GR deviation parameters which are a factor $2-6$ worse than the results using restricted priors on \LCDM parameters.}

\begin{document}
\maketitle
\flushbottom

\section{Introduction}

The Standard Cosmological model, also referred to as $\Lambda$CDM, is a successful model of the Universe, whose predictions have passed many precise observational tests in both the early and late-time universe \cite{2016PDU....12...56B}. 
Nonetheless, despite its match to a large set of cosmological measurements, the \LCDM still suffers from  important experimental and theoretical difficulties. 
On the experimental side, there are discrepancies between  independent measurements of the Hubble constant $H_0$ — the expansion rate of the Universe today — which is a fundamental parameter. 
The most cited tension is a $4.2\sigma$ discrepancy \cite{2017NatAs...1E.169F} between $H_0$ = $66.93 \pm 0.62 \, \hu$ inferred by the Planck collaboration from Cosmic Microwave Background (CMB) \cite{2016A&A...594A..13P}, and the $H_0 = 73.5 \pm 1.4 \hu$ \cite{2019ApJ...886L..27R,2019NatRP...2...10R} measured from Type Ia Supernovae. 
On the theoretical side, the nature of the two largest energy contributions, Dark Energy (DE) and Cold Dark Matter (CDM), still remains unknown \cite{2011PhRvD..83b3005Z}. 
While there are reasons to investigate CDM as composed of weakly interacting massive particles, few hints exist concerning the physical origin of DE. 
Alternative theories of gravity have been formulated to provide an explanation of the nature of DE on cosmological scales, and also perhaps solve the $H_0$ tension, see e.g.~\cite{2018FrASS...5...44E,2019JCAP...07..024B,Deffayet:2013lga} for a review of different modified gravity theories. In these theories the dynamics of perturbations are modified, and hence the propagation of tensor perturbations, namely GWs, differs from that of General Relativity (GR) \cite{2013PhRvD..87l3532A,2019PhRvD..99h3504L,2019PhRvD.100l3501G}. 

GWs offer a unique opportunity to test both cosmology and modified theories of gravity. 
Indeed, using compact binary systems as standard sirens \cite{1986Natur.323..310S}, it is possible to infer directly their luminosity distance $d_L^{\rm{GW}}$ \cite{1986Natur.323..310S,2005ApJ...629...15H,2008PhRvD..77d3512M,2009LRR....12....2S,2010ApJ...725..496N} without the use of a cosmological ladder.
In General Relativity, GWs are detected with an amplitude inversely proportional to the standard, photon luminosity distance of the source $d_L^{\rm EM}$ (here EM stands for electromagnetic).
In this paper, we consider modified GW propagation equations having a different friction term relative to GR, and consequently a GW luminosity distance $d_L^{\rm GW} \neq d_L^{\rm EM}$.
If provided with an observed EM counterpart and its redshift estimation, GWs can be used to measure both cosmological parameters $(H_0,\Omega_{m,0})$ and the parameters describing the modified GW friction term. 
Assuming GR is the correct theory of gravity, $H_0$ has been measured using the GW170817 hosting galaxy identification in \cite{2017Natur.551...85A}; and for standard sirens without an observed EM counterpart, the same has been done using galaxy surveys~\cite{2019ApJ...871L..13F,2019arXiv190806060T,2019ApJ...876L...7S,2020ApJ...900L..33P,2020ApJ...896L..44A} or methods studying the clustering of GW signals~\cite{Mukherjee:2019wfw,Mukherjee:2019wcg,Mukherjee:2020hyn}. In the context of modified gravity, GW170817 has been used to probe several parametrisations of the modified GW friction term~\cite{Pardo:2018ipy,2019arXiv190703150F,2018JCAP...03..021L,2019PhRvD.100d4041D,2020JCAP...03..015B,2018JCAP...07..048P,2019PhRvD..99h3504L,2019PhRvL.123a1102A,2020PhRvD.102d4009M}, 
however those constraints are not very stringent due to the low-redshift of the event. 
Moreover, if $(H_0,\Omega_{m,0})$ were left to vary, it was not possible to constrain the GR deviation parameters and $(H_0,\Omega_{m,0})$ together. This is due to the fact that the GW friction parameter and $H_0$ are strongly degenerate at low-redshift \cite{2020PhRvD.102d4009M}.

In this paper we use the GW events GW170817 \cite{2017PhRvL.119p1101A,2019PhRvX...9a1001A} and GW190521 \cite{2020PhRvL.125j1102A,Abbott_2020}  to constrain theories of gravity with a modified fiction term.
GW190521 is the furthest detection achieved by the LIGO \cite{2015CQGra..32g4001L} and Virgo \cite{2015CQGra..32b4001A} detectors during their third observing run O3 \cite{Tse:2019wcy,Acernese:2019sbr} at a luminosity distance of $\sim 4-5$ Gpc. The detection is associated with the merger of two black holes producing an intermediate mass black hole.
\textit{ZTF19abanrhr} \cite{2020PhRvL.124y1102G} is the tentative EM counterpart associated with GW190521, produced by the merger of two black holes in an AGN disk, at redshift $z=0.438$.
Assuming that GW190521 and \textit{ZTF19abanrhr} are associated with the same astrophysical source, we constrain modified theories of gravity. We do not report any deviation from GR.

The paper is organized as follows: in Sec.~\ref{sec:2} we discuss the inferencial and data analysis frameworks; in Sec.~\ref{sec:3} we introduce the models of modified gravity and their effects; in Sec.~\ref{sec:4} we present our results; and in Sec.~\ref{sec:5} we discuss them in light of previous results. Finally in Sec.~\ref{sec:6} we draw our conclusions.

\section{Data analysis method \label{sec:2}}

Given a set of cosmological parameters $\vec{\Lambda}=\{H_0,\Omega_{m,0}\}$ and GR deviation parameters $\vec{\alpha}$, the likelihood of having the observed GW data $x_{\rm GW}$ and EM data $x_{\rm EM}$
can be written as
%
%
\begin{equation}
    p(x_{\rm GW},x_{\rm EM}|\vec{\Lambda},\vec{\alpha})= \frac{1}{\beta(\vec{\Lambda},\vec{\alpha})} \int dz\, d\vec{\theta}  \; p(x_{\rm GW}|\vec{\theta})   p(x_{\rm EM}|z) p(\vec \theta|z, \vec \Lambda,\vec{\alpha}) p(z|\vec{\Lambda},\vec \alpha). 
    \label{eq:rough_likeli}
\end{equation}
Here $z$ is the redshift, and $\vec{\theta}$ are the GW parameters which include the masses in the detector frame and the GW luminosity distance $d_L^{\rm GW}$. The factor of $\beta(\vec{\Lambda},\vec{\alpha})$ encodes the {\it selection effects} (see below).  
The GW likelihood  is denoted by $p(x_{\rm GW}|\vec{\theta})$, whereas $p(x_{\rm EM}|z)$ is the EM counterpart observation likelihood, and $p(z|\vec{\Lambda},\vec{\alpha})$ is a uniform in comoving volume-time prior, independent of the GR deviation parameters. 
The term $p(\vec \theta|z, \vec \Lambda,\vec{\alpha})$ encodes the probability of having a set of GW parameters $\vec{\theta}$ from a given set of $(z,\vec{\Lambda}, \vec \alpha)$. 
Some of the GW parameters, such as the GW luminosity distance $d^{\rm GW}_{L}$ are functions of $(z,\vec{\Lambda},\vec{\alpha})$ and for these parameters we can write $p(d^{\rm GW}_L|z,\vec \Lambda,\vec{\alpha})=\delta(d^{\rm GW}_{L} - d^{\rm GW}_{L}(z,\vec \Lambda,\vec{\alpha}))$.
For all of the remaining GWs parameters we choose priors independent of $(z, \vec \Lambda,\vec{\alpha})$  that match the default priors used by LIGO and Virgo. 

Regarding the distribution of masses in the {\it detector} frame, we assume a uniform prior following \cite{Mukherjee:2020kki,Chen:2020gek}. 
Given the current uncertainties in the population mass models in the source frame, this is a reasonable assumption \cite{2020arXiv200905472F}. 
However, in a more accurate analysis, and in particular when combining a large number of events, physically motivated priors should be set on the source frame masses. 
Indeed, the choice of source frame prior can be used for cosmological inference as shown in \citep{2012PhRvD..85b3535T,2019ApJ...883L..42F}.
We therefore use uniform priors on masses in the detector frame, compatible to the ones used by LVC when computing the detectable fraction factor.
This allows us to write Eq.~(\ref{eq:rough_likeli}) as 
\begin{eqnarray}
    p(x_{\rm GW},x_{\rm EM}|\vec{\Lambda},\vec{\alpha})= \frac{1}{\beta(\vec{\Lambda},\vec{\alpha})} \int dz\, p(x_{\rm GW}|d_{L}^{\rm GW}(z,\vec \Lambda,\vec \alpha)) p(x_{\rm EM}|z) p(z|\vec{\Lambda},\vec \alpha). 
    \label{eq:rough_likeli2}
\end{eqnarray}

The factor $\beta$ in Eqs.~(\ref{eq:rough_likeli}-\ref{eq:rough_likeli2}),  encodes the selection effects \cite{2019ApJ...871L..13F,2019PhRvD..99h3504L,2019MNRAS.486.1086M}, and takes into account that for some choice of the population parameters, events can be either easier or more difficult to detect.
In order to take selection effects into account, we compute the detectable fraction
\begin{equation}
    \beta(\vec{\Lambda},\vec{\alpha})= \int dz   P_{\rm det}^{\rm GW}(d^{\rm GW}_{L} (z,\vec \Lambda,\vec{\alpha})) P^{\rm EM}_{\rm det}(z) p(z|\vec{\Lambda}),
\end{equation}
where $P_{\rm det}^{\rm GW}(d^{\rm GW}_{L})$ is the probability of detecting a binary at a distance $d^{\rm GW}_{L}$.
In this paper we make the assumption that the AGN flare can {\it always} be detected following a GW like GW190521, so that $P_{\rm EM}^{\rm det}(z)=1$. The same assumption was made for GW170817, as the detection horizon for short $\gamma$-ray burst and Kilonova transients during O2 was significantly higher than the detection horizon of LIGO and Virgo for GW \cite{2018Natur.562..545C,2019ApJ...871L..13F}.
The detection probability for GW170817 was computed using software injections in simulated LIGO and Virgo data from sensitivities representative of O2 provided with the GW170817 data distribution\footnote{https://dcc.ligo.org/LIGO-P1800061/public}, whilst for GW190521 we have used sensitivities representative of O3a provided with the event data distribution\footnote{https://dcc.ligo.org/LIGO-P2000158/public}.

In order to compute the GW likelihood term as a function of the GW luminosity distance, we renormalize the posterior samples provided by the LIGO and Virgo collaboration by a quadratic prior on luminosity distance (used to produce the analysis).
For GW170817 we use the the high spin posterior samples of \texttt{IMRPhenomPv2\_NRTidal} \cite{2019PhRvX...9a1001A}, 
while for GW190521 we use the posterior samples provided by the three waveform posteriors presented in \cite{Abbott_2020}: \texttt{NRSur} \cite{2019PhRvR...1c3015V}, \texttt{IMRPhenom} \cite{2014LRR....17....2B,2020PhRvD.101b4056K} and \texttt{SEOBNR} \cite{2020PhRvD.102d4055O}.
We calculate the \textit{ZTF19abanrhr} line-of-sight posterior distribution on the GW luminosity distance by selecting all the posterior samples within $0.0045 \,  {\rm deg}^2$ from the AGN location and then by marginalizing the selected posterior samples over all the variables with the exception of the luminosity distance. The solid angle is chosen in such a way to be small enough around the ZTF counterpart but large enough to contain a reasonable number of samples for fitting the marginal distance posterior. Several values of the solid angle have been explored in order to check the validity of our fit, see Appendix \ref{eq:appB} for more details.

Regarding the redshift estimation of GW170817, we assume a Gaussian distribution centered at the value of $z=%
$ 0.01003 with standard deviation 0.0005. The uncertainty on the GW170817 redshift is mostly due to the corrections related to peculiar velocities of NGC4993 \cite{2019PhRvX...9a1001A,Mukherjee:2019qmm}. As concerns GW190521, we fix the redshift to the value of $0.438$ \cite{2020PhRvL.124y1102G}, neglecting uncertainties on redshift and peculiar velocity corrections. This should be a good approximation for sources at these larger redshifts.

\begin{figure}[h!]
    \centering
    \includegraphics[scale=0.5]{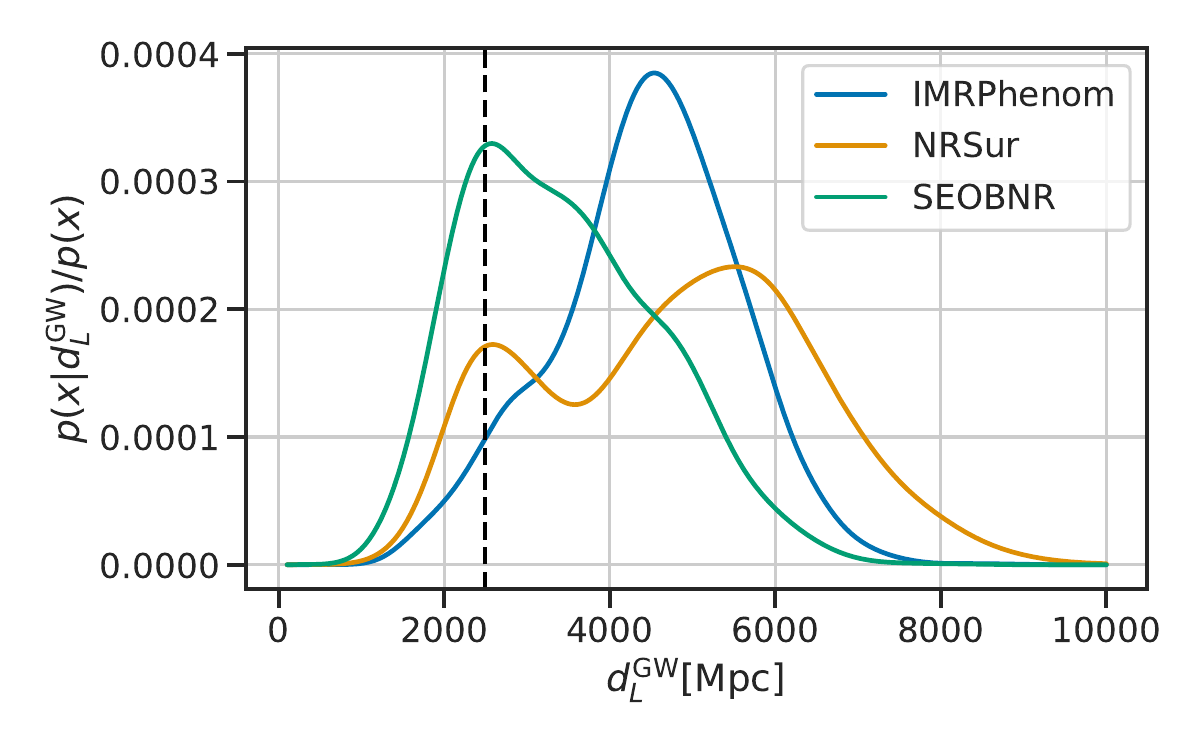}
    \caption{Line-of-sight luminosity distance posteriors for the three  waveform models for GW190521, renormalized by a quadratic prior in luminosity distance. The vertical dashed line indicates the \LCDM luminosity distance at $z=0.438$. Cosmological parameters are set to Planck's values \cite{2018arXiv180706209P}.}
    \label{fig:los_compare}
\end{figure}

Fig.~\ref{fig:los_compare} shows the waveforms posteriors on GW luminosity distance renormalized with a quadratic prior on the GW luminosity distance. The three waveforms predict a different luminosity distance, although all three show a peak around the \LCDM luminosity distance computed assuming no GR deviations. (black dashed line)

\section{GW propagation model \label{sec:3}}

In an expanding \LCDM universe, and taking into account the modified GW friction term, the propagation equation for the GW polarizations $h$ (we drop the polarization index for simplicity) is given by \cite{2018PhRvD..97j4066B}
\begin{equation}
    h'' + [2+\alpha_M (\eta)]\frac{a'}{a} h' +k^2 c^2 h =0, 
    \label{eq:gem}
\end{equation}
where $c$ is the speed of light, $\alpha_M (\eta)$ is the GW friction parameter, $a$ is the scale factor, $k$ the comoving wavenumber, and derivatives are with respect to the conformal time $\eta$. 
It follows that the luminosity distance of the detected GWs in the detector frame is given by
\begin{equation}
    d_L^{\rm GW}(z)=d_L^{\rm EM}(z) \exp \left[ \frac{1}{2} \int_0^z \frac{\alpha_M(z')}{1+z'}dz' \right],
    \label{eq:GWlum_gen}
\end{equation}
where $z$ is the cosmological redshift. 
The form of $\alpha_M$ is determined by a given theory of gravity. In General Relativity $\alpha_M(z)=0$ for all $z$, whereas in modified gravity theories it is generally non-vanishing. In this sense
 GW190521 is particularly interesting as its higher redshift $z \sim 0.44$ (relative to GW170817) enables us to probe $d_L^{GW}(z)$ (and hence $\alpha_M(z)$) at higher $z$.

We now consider three different parametrisations of $d_L^{\rm GW}(z)$ or $\alpha_M(z)$ which have been proposed in the literature. Each corresponds to models that, amongst other reasons, attempt to explain the late-time accelerated expansion of the Universe (without the need for a cosmological constant) and to provide an explanation for the different behaviour of gravity at different scales and epochs which might then explain the discrepancies observed between the Hubble constant at the epoch of the CMB and in the local universe (today). The models we describe below modify gravity in different ways, see \cite{2019JCAP...07..024B} for a detailed discussion, and they also introduce a different relation for the GW luminosity distance. These models are usually well constrained at the level of the CMB, see Sec.~\ref{sec:5} for a discussion on the different constraints in comparison with the ones found in this paper.

\subsection{$\Xi$-parametrization of $d_L^{\rm GW}$}

Scalar-tensor theories of gravity, in which an additional scalar field couples the spin-2 graviton, have long been studied as alternative theories of gravity. 
Several classes of increasing complexity have been developed, including the Brans-Dicke~\cite{Brans:1961sx};  Horndeski~\cite{1974IJTP...10..363H,Deffayet:2011gz,Kobayashi:2011nu}, beyond-Horndeski~\cite{Gleyzes:2014dya}, and DHOST~\cite{Langlois:2015cwa} theories.
As discussed in \cite{Belgacem:2019pkk} and first proposed in \cite{2018PhRvD..97j4066B}, for some of these theories (and also for others including the RR and RT models \cite{Maggiore:2013mea,Maggiore:2014sia}) the GW luminosity distance is well parametrised by
\begin{equation}
    d_L^{\rm GW} = d_L^{\rm EM}  \left[\Xi + \frac{1-\Xi}{(1+z)^n}\right],
\end{equation}
where $\Xi,n>0$.
GR is recovered when $\Xi=1$, and more generally when $\Xi \neq 1$ as $z\rightarrow 0$.
As no external constraint on $\Xi$ is available from previous measurements, we probe a log-uniform prior on $\Xi=1$ spanning in the range $[0.01,100]$.
The prior on the stiffness parameter $n$ is
similarly chosen to be uniform within the range $[1,10]$.

\subsection{Extra dimensions}

Some modified gravity models, such as DGP gravity \cite{2000PhLB..485..208D} and  some models of quantum gravity \cite{2019PhLB..79835000C}, have their origins in extra dimensional space-times: they are characterised by an additional length scale $R_c$, beyond which gravity deviates from GR. It follows from flux conservation that $d_L^{\rm GW}$ is modified on these scales, and a parameterisation proposed in \cite{Deffayet:2007kf}  for non-compactified extra dimensions is
\begin{equation}
    d_L^{\rm GW} = \left[1+\left(\frac{d_L^{\rm EM}}{R_c}\right)^n \right] ^{\frac{D-2}{2n}},
    \label{eq:dpg}
\end{equation}
where the parameter $n$ encodes the stiffness of the transition and $D$ the number of space-time dimensions.  
Here we assume that at the cosmological scales we are probing with these GW events, $R_c \ll d_L^{\rm EM}$. 
In that case Eq.~(\ref{eq:dpg}) reduces to the simpler form
\begin{equation}
    d_L^{\rm GW} = (d_L^{\rm EM})^{\frac{D-2}{2}}.
\end{equation}
In this work we take a uniform prior around the GR expected value $D \in [3,7]$.  (For other parametrisations and constraints from GWs on extra-dimensional theories, see \cite{Pardo:2018ipy}.)

\begin{table*}[h!t]
\caption{Upper limits and constraints at 95\% CL for several parameters describing three models of alternative theories of GR considered. We report upper limits and constraints for each of the waveform model (see Fig.~\ref{fig:los_compare}) and  different choices of priors on  \LCDM parameters. ``Planck's''  refers to gaussian priors on $H_0 =\mathcal{N}(\mu =67.66,\sigma=0.42) \hu$ and $\Omega_{m,0}=\mathcal{N}(\mu =0.311,\sigma=0.056)$\cite{2018arXiv180706209P} and 
``Wide'' refers to uniform priors in the range $H_0 \in [20,300] \hu$ and $\Omega_{m,0}\in [0.2,1.0]$. \label{tab:tab1}}
\begin{center}
\begin{tabular}{|ccc|cc|cc|}

    \hline
    Model & \multicolumn{6}{c|}{\textbf{$c_M$-parametrization}}\\
    \hline
    Waveform & \multicolumn{2}{c|}{\texttt{NRSur}} & \multicolumn{2}{c|}{\texttt{IMRPhenom}} & \multicolumn{2}{c|}{\texttt{SEONBR}} \\
    \hline
    Prior & Wide & Planck & Wide & Planck & Wide & Planck  \\
    $H_0\rm{[\hu]}$
&$80^{+67}_{-17}$ & $67.7^{+0.8}_{-0.8}$
&$79^{+60}_{-16}$ & $67.7^{+0.8}_{-0.8}$
&$81^{+68}_{-18}$ & $67.7^{+0.8}_{-0.8}$ \\
$\Omega_{m,0}$ 
&$0.6^{+0.4}_{-0.4}$ & $0.311^{+0.011}_{-0.011}$
&$0.6^{+0.4}_{-0.4}$ & $0.311^{+0.011}_{-0.011}$
&$0.6^{+0.3}_{-0.4}$ & $0.311^{+0.011}_{-0.011}$ \\
$c_M$ 
& $<13.4$ & $<7.5$
& $<11.7$ & $<6.3$
& $<10.8$ & $<5.7$ \\

\hline
    Model & \multicolumn{6}{c|}{\textbf{Extra dimension}}\\
    \hline
    Waveform & \multicolumn{2}{c|}{\texttt{NRSur}} & \multicolumn{2}{c|}{\texttt{IMRPhenom}} & \multicolumn{2}{c|}{\texttt{SEONBR}} \\
    \hline
     Prior & Wide & Planck & Wide & Planck & Wide & Planck  \\
    $H_0 \rm{[\hu]}$
&$167^{+113}_{-83}$ & $67.7^{+0.8}_{-0.8}$
&$152^{+120}_{-65}$ & $67.7^{+0.8}_{-0.8}$
&$141^{+130}_{-74}$ & $67.7^{+0.8}_{-0.8}$ \\
$\Omega_{m,0}$
&$0.6^{0.4}_{-0.4}$ & $0.311^{0.011}_{-0.011}$
&$0.6^{0.4}_{-0.4}$ & $0.311^{0.010}_{-0.011}$
&$0.6^{0.4}_{-0.4}$ & $0.311^{0.011}_{-0.010}$ \\
$D$
&$4.5^{+0.3}_{-0.4}$ & $3.99^{+0.08}_{-0.09}$
&$4.5^{+0.3}_{-0.3}$ & $4.01^{+0.10}_{-0.10}$
&$4.4^{+0.3}_{-0.4}$ & $4.00^{+0.09}_{-0.10}$ \\

\hline
    Model & \multicolumn{6}{c|}{\textbf{$\Xi$-parametrization}}\\
    \hline
    Waveform & \multicolumn{2}{c|}{\texttt{NRSur}} & \multicolumn{2}{c|}{\texttt{IMRPhenom}} & \multicolumn{2}{c|}{\texttt{SEONBR}} \\
    \hline
     Prior & Wide & Planck & Wide & Planck & Wide & Planck  \\
    $H_0\rm{[\hu]}$
&$93^{+148}_{-27}$ & $67.7^{+0.8}_{-0.8}$
&$90^{+126}_{-25}$ & $67.7^{+0.8}_{-0.8}$
&$87^{+119}_{-23}$ & $67.7^{+0.8}_{-0.8}$ \\
$\Omega_{m,0}$
&$0.6^{0.4}_{-0.4}$ & $0.311^{0.011}_{-0.010}$
&$0.6^{0.4}_{-0.4}$ & $0.311^{0.011}_{-0.011}$
&$0.6^{0.4}_{-0.4}$ & $0.311^{0.010}_{-0.011}$ \\
$\Xi$
& $<10$ & $<4.4$
& $<8.2$ & $<3.6$
& $<7.1$ & $<2.9$ \\
$n$
&$6^{+4}_{-5}$ & $4^{+5}_{-4}$
&$6^{+4}_{-5}$ & $5^{+5}_{-4}$
&$6^{+4}_{-5}$ & $5^{+5}_{-4}$ \\

\hline

\end{tabular}
\end{center}
\end{table*}

\subsection{$c_M$-parametrization}

Rather than parametrising $d_L^{\rm GW}(z)$ as above, another approach advocated in the literature is to parametrize the friction term $\alpha_M(z)$. In particular, in \cite{2019PhRvD..99h3504L}, the authors propose
\begin{equation}
    \alpha_M(z)= c_M \frac{\Omega_{\Lambda}(z)}{\Omega_{\Lambda}(0)},
\end{equation}
where $c_M$ is a constant, and $\Omega_{\Lambda}(z)$ is the fractional dark energy density. (GR is recovered when $c_M=0$.)
Indeed for modified gravity models trying to explain dark energy, it is reasonable to assume that $\alpha_M$ is linked to the evolution of the dark energy content of the universe.
%
Substituting in Eq.~(\ref{eq:GWlum_gen}) gives \cite{2019PhRvD..99h3504L}
\begin{equation}
d_L^{\rm GW} = d_L^{\rm {EM}} {\rm{exp}} \left[\frac{c_M}{2\Omega_{\Lambda,0}} \ln \frac{1+z}{\Omega_{m,0}(1+z)^3+\Omega_{\Lambda,0}} \right]
\end{equation}
In this paper we take a uniform prior for $c_m \in [0,150]$. We intentionally  exclude negative values of $c_M$ as for very high redshift, $d_L^{\rm GW}$  can decrease with redshift. 
This is clearly a not physical situation as it would correspond to the possibility of detecting very high redshift sources with an infinite SNR and it causes also problems for the computation of the selection effect.

\section{Results \label{sec:4}}

As a first check, we estimate  $H_0$ and $\Omega_{m,0}$ in the GR limits (no GW friction term). We obtain estimates of these two parameters consistent with previous studies in \cite{Mukherjee:2020kki,Chen:2020gek}, see App.~\ref{eq:appA} for more details.

We then perform two runs with different priors for the \LCDM parameters $H_0$ and $\Omega_{m,0}$.
In the first run, we fix the Gaussian priors on $H_0 =\mathcal{N}(\mu =67.66,\sigma=0.42) \hu$ and $\Omega_{m,0}=\mathcal{N}(\mu =0.311,\sigma=0.056)$ to the measured Planck's values of \cite{2018arXiv180706209P}, where $\mathcal{N}(\mu,\sigma)$ is a gaussian distribution with mean $\mu$ and standard deviation $\sigma$. In the second run we take a uniform prior on both $H_0 \in [20,300] \hu$, and $\Omega_{m,0}\in [0.2,1.0]$.

\begin{figure}
    \centering
    \includegraphics[scale=0.45]{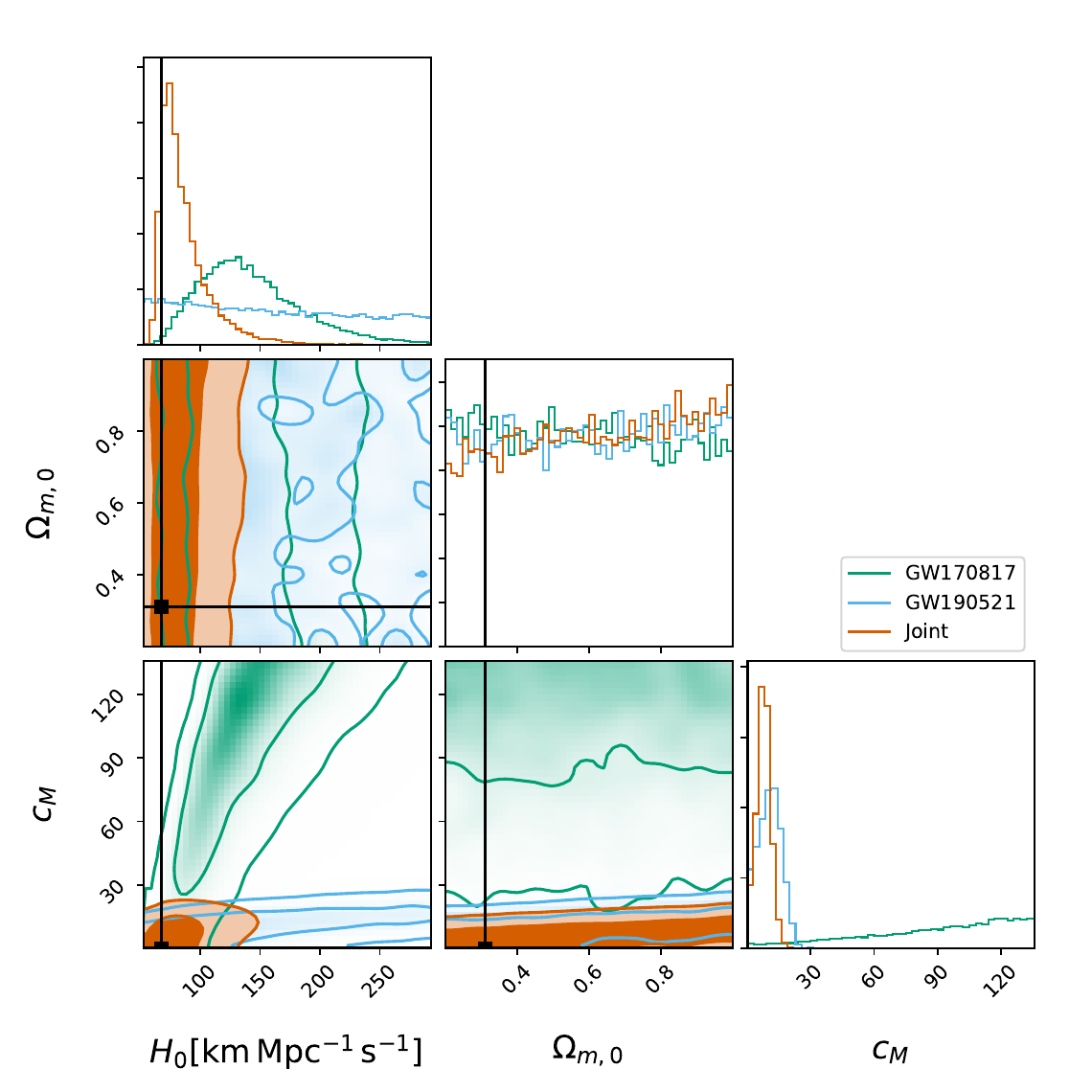}
    \caption{Posterior distributions obtained GW170817 (green), GW190521 \texttt{NRSur} waveform (blue) and the joint analysis (red) for wide priors in the case of the $c_M$-parametrization. The contours show the $68\%$ and $95\%$ CL intervals. The black lines indicates the GR-\LCDM parameters measured by Planck\cite{2018arXiv180706209P}.}
    \label{fig:pmass}
\end{figure}

\begin{figure}
    \centering
    \includegraphics[scale=0.45]{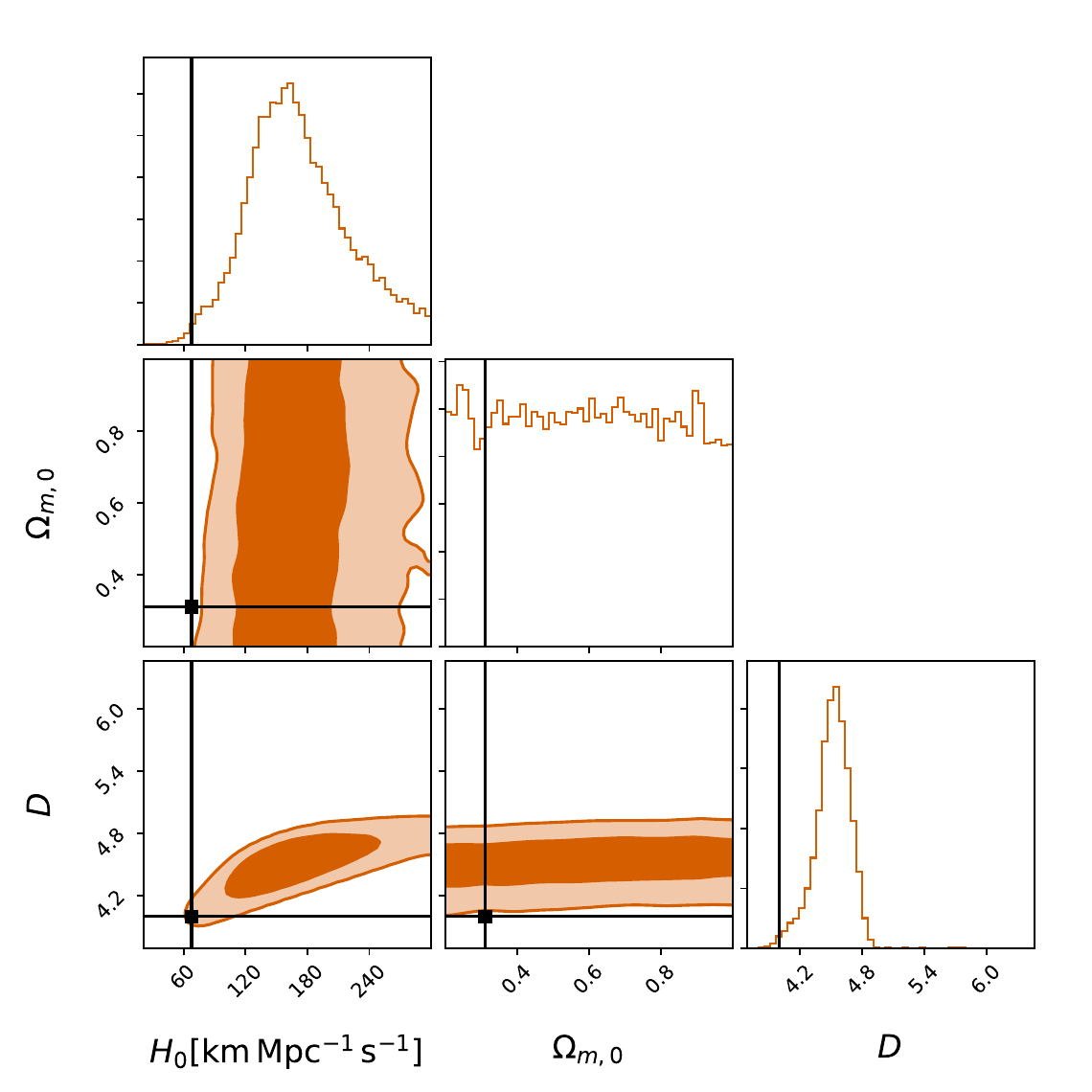}
    \caption{Posterior distributions for the GW170817-GW190521 (\texttt{NRSur}) analysis (red) for wide priors in the case of the Extra dimensions gravity. The contours show the $68\%$ and $95\%$ CL intervals. The black lines indicates the GR-\LCDM parameters measured by Planck\cite{2018arXiv180706209P}.}
    \label{fig:D_dimensions}
\end{figure}

\begin{figure}
    \centering
    \includegraphics[scale=0.45]{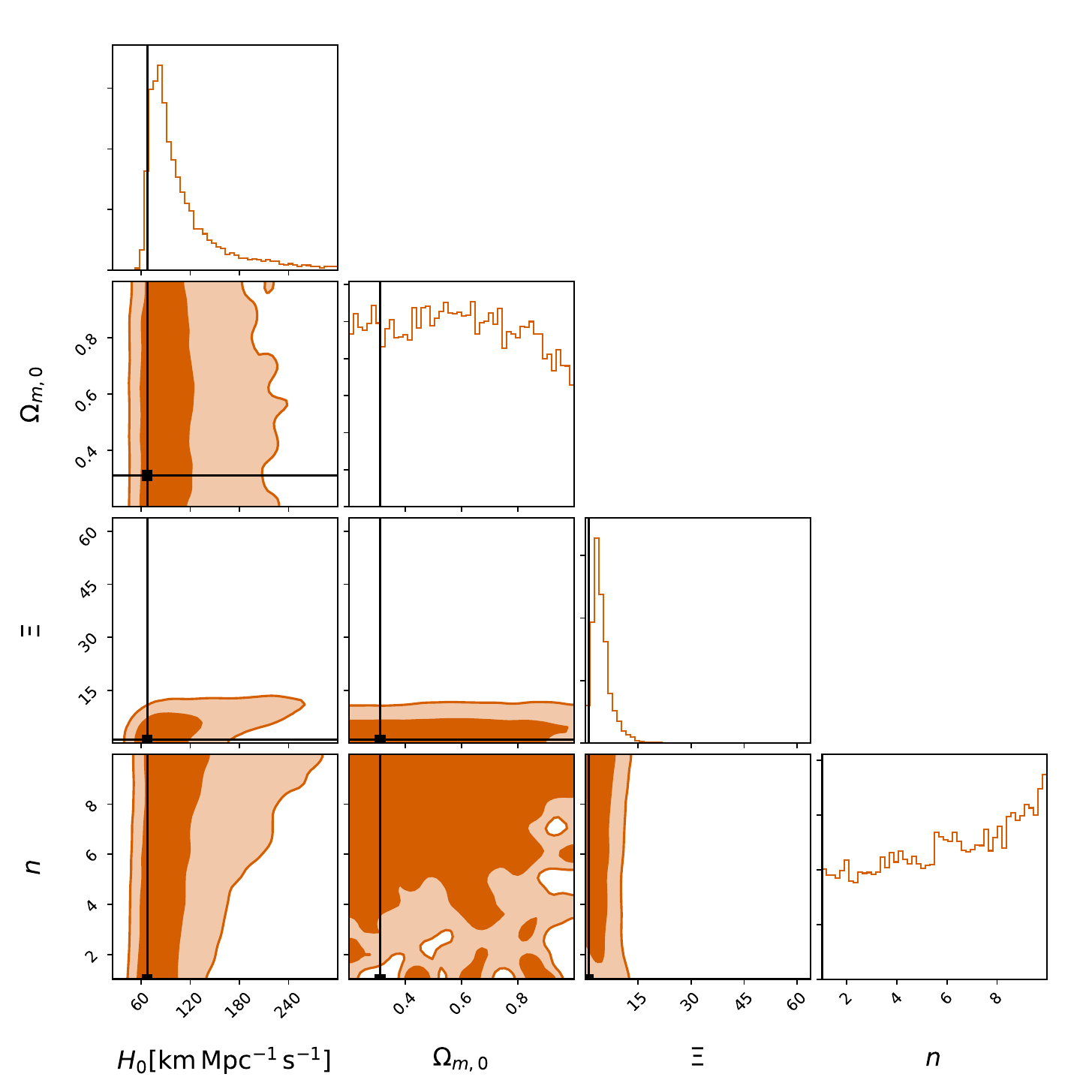}
    \caption{Posterior distributions for the GW170817-GW190521 (\texttt{NRSur}) analysis (red) for wide priors in the case of the $\Xi$-parametrization. The contours show the $68\%$ and $95\%$ CL intervals. The black lines indicates the GR-\LCDM parameters measured by Planck\cite{2018arXiv180706209P}.}
    \label{fig:RR_model}
\end{figure}

We now discuss our results in the context of different modified gravity theories. First of all, for all parametrizations of the GR deviations, we find that in the case of wide priors, $H_0$ can be constrained (with high uncertainty if compared to the Planck's values) while $\Omega_{m,0}$ can not be constrained. See Figs.~\ref{fig:pmass}-\ref{fig:D_dimensions}-\ref{fig:RR_model}. 

The results for the different theories and priors are given in Tab.~\ref{tab:tab1}.
In the most general case with uniform priors on cosmological parameters, we find that GW170817 and GW190521 with their EM counterparts can provide a joint constraint on $H_0$ and on the GR deviation parameters.
The effect of combining these two events is shown in Fig.~\ref{fig:pmass} for the $c_M$-parametrization.
In principle, neither of these two events separately can provide a joint constraint on $H_0$ and $c_M$, since these two parameters are degenerate with each other. 
However, as can be seen from Fig.~\ref{fig:pmass}, the posteriors have different shapes due to the different redshift of the two events.
For GW170817, a small variation of $H_0$ can be compensated by a large variation of $c_M$ while for GW190521 the contrary is true. This can also be seen from Eq.~(\ref{eq:GWlum_gen}).
From the joint analysis of the two events, GW170817 provides a good estimate of $H_0$, while GW190521 provides a good constraint on $c_M$.
By combining their posteriors it is possible to provide a constraints on both parameters, see Tab.~\ref{tab:tab1} for the final values.

For the three parametrisations considered in Sec.~\ref{sec:3}, all the runs with wide \LCDM priors are compatible in $1-2\sigma$ confidence Level (CL) with GR and cannot reach the precision that would be needed to solve the $H_0$ tension. The larger deviation is usually given by the posterior associated to the waveform approximant \texttt{IMRPhenom} that has lower support for a source at redshift $\sim 0.44$ in GR, see Fig.~\ref{fig:los_compare}.
The constraints and upper limits (ULs) that we provide below are given at 95\% CL.
We find that for all the waveform models the $c_M$ UL is $<14$ while $H_0$ can be constrained to $H_0 \sim 80^{+67}_{-17} \hu$.
For the Planck prior we find that the upper limit on $c_M$ improves by a factor of two with respect to wide priors, and we obtain posteriors on $H_0$ and $\Omega_{m,0}$ consistent with measured values of \cite{2018arXiv180706209P}.

Fig.~\ref{fig:D_dimensions} shows our results for the extra-dimensions model with wide priors.
Here $H_0$ is poorly constrained to a value of  $H_0\sim 167^{+113}_{-83} \hu$ while the number of space-time dimensions is measured with an precision of $20\%$ and it is compatible with 4 space-time dimensions at $\sim 2.1\sigma$. This deviation is mostly due to the GW190521 \texttt{NRSur} and \texttt{IMRPhenom} posterior supporting higher redshift values.
If we assume Planck's priors on \LCDM parameters, we improve the precision of $D$ measurement to an precision of $5\%$ still compatible with GR within $<1\sigma$ CL.

Finally, Fig.~\ref{fig:RR_model} shows results for the $\Xi-$parametrization. 
In the case of wide \LCDM priors we find that $H_0$ can be constrained to $ H_0\sim 98^{+148}_{-27} \hu$ while the upper limit on $\Xi$ is found to be $<10$ at 95\% CL. When assuming \LCDM Planck's prior, we improve the constraint on $\Xi$ by a factor of two and recover the Planck's priors for the \LCDM parameters.
With either wide or Planck's priors, we find that the stiffness coefficient $n$ cannot be constrained and returns the prior distribution. This is mostly because the redshift of GW190521 is not comparable with the cosmological redshifts at which this parameter could be constrained \cite{2018JCAP...03..002B}.

\section{Discussion \label{sec:5}}

The inclusion of GW190521 significantly enhances our ability to constrain modified theories of gravity with respect to previous studies. As mentioned before, when we use wide priors on \LCDM parameters, the posterior we find on $\Omega_{m,0}$ is uninformative.

For extra-dimension gravity, the previous constraints of spacetime dimensions from GW170817 fixing a Planck's prior had an precision of $\sim 5\%$ (at 1$\sigma$ CL) \cite{2018JCAP...07..048P} and $H_0$ was not constrained with this modified gravity model. By including GW190521, we improve the precision of a factor of $\sim 2$ reaching an precision of $2.5\%$ (at $1\sigma$ CL).
On the other hand, when using wide priors, we obtain a precision on $D$ of $\sim 20\%$ even though in this case the width of the $H_0$ posterior is very large $\sim 200 \hu$ and due to the high support of GW190521 posterior to higher distances, higher values of spacetime dimensions are preferred.
Note that in this work we have assumed that extra dimension modifications will appear as soon as we enter into the Hubble flow. However, some modifications of gravity might appear at higher scales such as 4~Gpc\cite{2000PhLB..485..208D}. In this case the number of spacetime dimensions would be more poorly constrained.

For the $c_M$ parametrization, previous studies \cite{2019PhRvD..99h3504L} on $c_M$ using GW170817 and Planck's priors on the \LCDM parameters, were reporting an uncertainty of $\sim \Delta c_M=70$. The results using wide priors were not constraining on $c_M$ in the range $[-150,150]$ and only $H_0$ was measured with an uncertainty of $\sim 80 \hu$ (at $1\sigma$) CL \cite{2019PhRvD..99h3504L}. 
The addition of GW190521 brought a non-negligible improvement, in the case of Planck's priors: the UL of $c_M$ improves by a factor of $20$. In the case of the wide priors we are able to provide a measurement of $H_0$ which is $\sim 2$ times better and constrain $c_M$ conjointly for the first time.
Note that even in the case of wide priors, our constraints on $c_M$ are better than those obtained in the case of Planck's \LCDM priors for GW170817 only in \cite{2019PhRvD..99h3504L}.
Even though the constraint from GW190521 on $c_M$ is significantly improved, we are not still at the level of cosmological motivated constraints that report $-0.62<c_M<1.35$ \cite{2019PhRvD..99j3502N}, so an improvement of an order of $10$ is still needed to reach this precision. This might be achieved by combining other $\sim 100$ events similar to GW190521 (time-demanding scenario) or by finding higher redshift GW events with associated EM counterparts.

For the $\Xi$-parametrization, previous constrains from this model were provided in terms of $n(1-\Xi)$ \cite{2018PhRvD..98b3510B} and were provided with an uncertainty of  $\sim 40$ at $1\sigma$ CL. However, this type of constraint was valid only at low redshift and it is obtained converting the fractional uncertainty obtained on $H_0$ for GW170817 to a fractional uncertainty on $n(1-\Xi)$  \cite{2018PhRvD..98b3510B}. 
Since we combine two events at different redshifts, we provide directly upper limits on $\Xi$ and cosmological parameters. In some non-local theories of gravity such as the RR and RT models $\Xi$ is expected to deviate from 1 at the order of 3-6\% values, while $n$ is expected to be $2.5-2.6$ \cite{2019JCAP...07..024B}. Our results cannot yet rule out these models of modified gravity.


Interestingly, in many modified gravity theories $\alpha_M$ is related to an effective running Planck's mass $M_{\rm eff}(z)$, or alternatively to an varying effective  Newton's constant $G_{\rm eff}(z)$. For these theories it can be shown that \cite{2018FrASS...5...44E}
%
\begin{equation}
\frac{d^{\rm GW} (z)}{d^{\rm EM}(z)} = \frac{M_{\rm eff}(0)}{M_{\rm eff}(z)}  =\sqrt{\frac{G_{\rm eff}(z)}{G_{\rm eff}(0)}} \equiv \sqrt{\hat G(z)}.
\end{equation}
If we assume that the $c_M$ and $\Xi$ parametrizations are representative of these theories, then we can also obtain a constraint on the value of the Planck's mass and the Newton constant at the redshift of GW170817 and GW190521.
Let us just report the constraints obtained from the \texttt{NRSur} waveform model as all of them reproduce consistent results as shown in Tab.~\ref{tab:tab1}.

For the $\Xi$-parametrization using restricted priors  we obtain  $ \hat G (z)=1.10^{+0.20}_{-0.12}$   at the redshift of GW170817 and $ \hat G (z)=3.5^{+6.0}_{-3.0}$ at the redshift of GW190521.
While when we use wide priors we obtain  $ \hat G (z)=1.33^{+1.86}_{-0.33}$  at the redshift of GW170817 and $ \hat G (z)=11.8^{+98.0}_{-11.0}$ at the redshift of GW190521.

Note that for the $c_M$ parametrization we can just provide an upper limit. The lower limit will be $1$ and is given by the condition on our priors $c_M>0$.
For the $c_M$-parametrization using restricted priors  we obtain $ 1 \leq \hat G (z)<1.08$  at the redshift of GW170817 and $1 \leq \hat G (z)<9.3$ at the redshift of GW190521.
While when we use wide priors we obtain   $1 \leq \hat G (z)< 1.14$   at the redshift of GW170817 and $1 \leq \hat G (z)<26$ at the redshift of GW190521.

We note that the results on the variation of the Planck's mass and Newton's constant are compatible between the $c_M$ and $\Xi$ prescriptions. We also note that this analysis improves the variation of the Newton Constant previously reported as $-1<\hat G<8$ in \cite{2020arXiv200312832V} analyzing GW170817.

\section{Conclusions \label{sec:6}}

In this paper we have presented new results and upper limits on \LCDM parameters and different models/parameterizations of GR deviations at cosmological scales using the two GW events GW170817 and GW190521.
With the addition of GW190521 we find that not only it is possible to jointly constrain $H_0$ and the GR deviation parameters, but it is also possible to significantly improve previous GWs-based constraints.
We found that the precision on the \LCDM parameters is not enough to solve the $H_0$ tension when we assume wide priors on the \LCDM parameters.


The ULs on GR modifications are improved by a factor of $2-10$ (depending on the model) with respect to previous studies, showing that the joint EM and GW detections of high redshift standard sirens can be used to constrain GR modifications.
Indeed, these results are not yet accurate as their cosmologically motivated limits such as those from CMB etc. In order to reach that result we would need an improvement of a factor of $10$.

GW190521 and its tentative EM counterpart offer a good opportunity for testing cosmology and recently also \cite{Mukherjee:2020kki,Chen:2020gek} also studied this event to infer the Hubble constant.
Currently, it is still unclear if ZTF19abanrhr is the actual EM counterpart of GW190521. For instance Ref.~\cite{Ashton:2020kyr} suggest that there is not strong statistical evidence for this association,  Ref.~\cite{Yang:2020yoc,Gayathri:2020coq} discuss the possibility that GW190521 was by a closer high eccentric binary and \cite{CalderonBustillo:2020srq} study GW190521 as the merger of two proca stars.  Nevertheless, the analysis presented in this paper shows how GW190521-like events can be informative (when supplied with an EM counterpart) in constraining modifications of gravity at cosmological scales. Indeed, with higher-redshift GW sources, in particular expected to be seen in the LISA band, we will be able to constrain gravity modifications at cosmological scales with a very high precision \cite{Corman:2020pyr}.

Future GW detections with their EM counterparts will improve
the constraints on GR deviations at cosmological scales, in particular in the case that these are high-redshift detections such as the one proposed for GW190521 and ZTF19abanrhr.
If confirmed the association of GW190521 with ZTF19abanrhr will provide unprecedented tests of GR.

In this paper we have presented new results and upper limits on \LCDM parameters and different models/parameterizations of GR deviations at cosmological scales using the two GW events GW170817 and GW190521.
With the addition of GW190521 we find that not only it is possible to jointly constrain $H_0$ and the GR deviation parameters, but it is also possible to significantly improve previous GWs-based constraints.
We found that the precision on the \LCDM parameters is not enough to solve the $H_0$ tension when we assume wide priors on the \LCDM parameters.

\section*{Acknowledgements}
The authors thank  J.~Baird and M.~Hendry for the useful comments during the development of this work.
SM is grateful to the Geneva cosmology group for highlighting an error in the preparation of the data needed for the analysis which was partially affecting the results.
SM is supported by the LabEx UnivEarthS (ANR-10-LABX-0023 and ANR-18-IDEX-0001), of the European Gravitational Observatory and of the Paris Center for Cosmological Physics.
LH is supported by the Swiss National Science Foundation Early Postdoc Mobility grant 181461. IMH is supported by the NSF Graduate Research Fellowship Program under grant DGE-17247915. CK is partially  supported   by  the Spanish MINECO   under
the grants SEV-2016-0588 and PGC2018-101858-B-I00, some of which include
ERDF  funds  from  the  European  Union. IFAE  is  partially funded by
the CERCA program of the Generalitat de Catalunya.

\appendix

\section{Extraction of the line-of-sight $d^{\rm GW}_{L}$ posterior \label{eq:appB}}
\begin{figure*}
    \centering
    \includegraphics[scale=0.3]{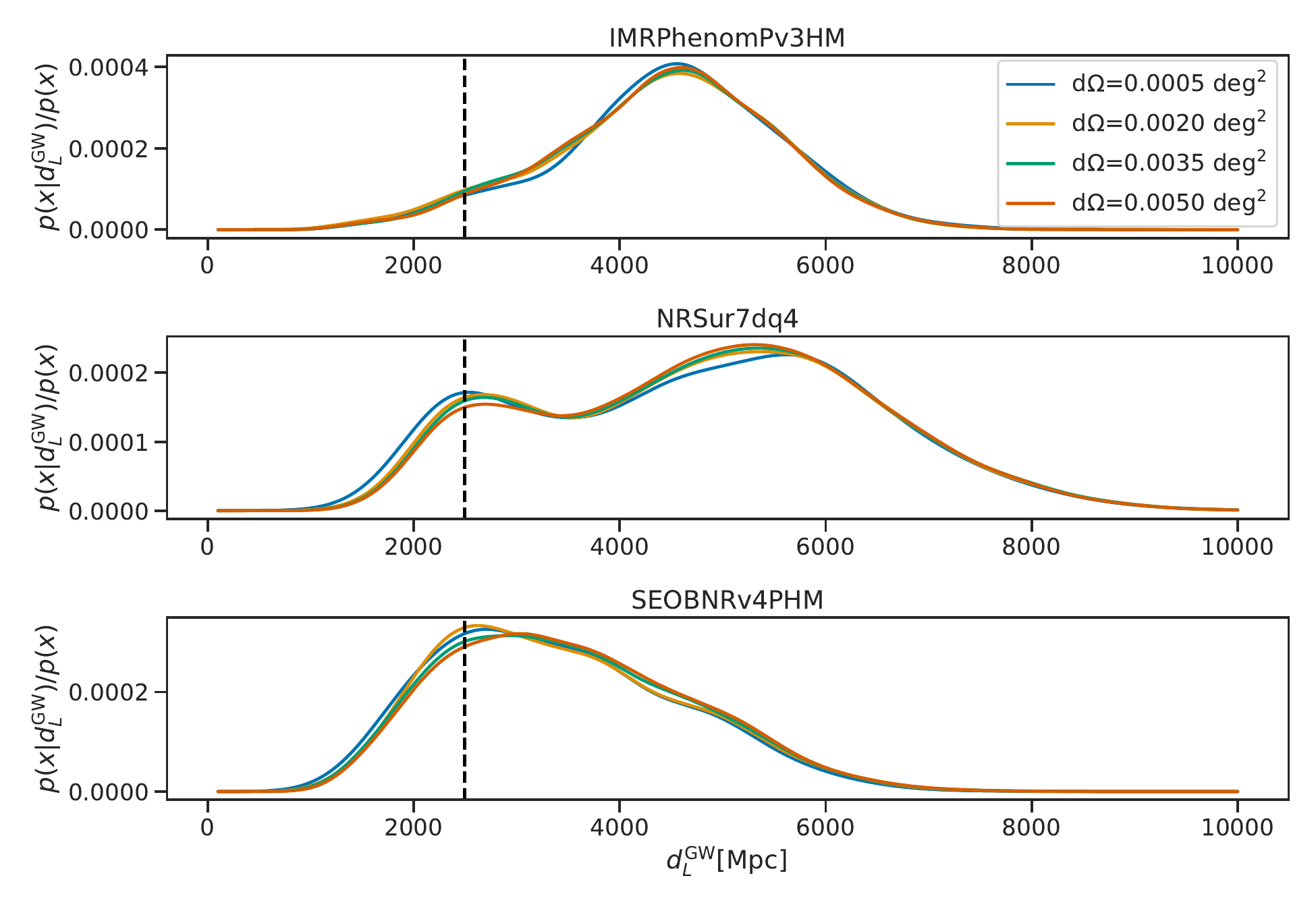}
    \caption{Variation of the line-of-sight luminosity distance posterior with respect to variations of the solid angle and renormalized by a quadratic prior in luminosity distance. The black vertical dashed line indicates the luminosity distance at $z = 0.438$ for a GR $\Lambda$CDM Universe with parameters set to Planck's values.}
    \label{fig:los_plot}
\end{figure*}

The line-of-sight $d^{\rm GW}_{L}$ luminosity distance posterior is calculated by keeping only the posterior samples that are within a solid angle corresponding to a certain sky area. In celestial coordinates the solid angle is given by:
\begin{equation}
    d \Omega= d \delta d \alpha \cos \delta .
\end{equation}
For the main analysis we chose this solid angle to be $0.0045 \, \rm{deg^{2}}$ (see Fig.~\ref{fig:los_compare}). Here we present how the line-of-sight posterior changes by varying the value of the solid angle from $ 0.0005 \, \rm{deg^{2}}$ up to $ 0.005 \, \rm{deg^{2}}$. In Fig.~\ref{fig:los_plot} we see the variation of the posterior while changing the solid angle. The number of posterior samples varied from around 1000 samples to 4500 samples with the aforementioned changing of the solid angle. However, the posterior does not change significantly, indicating that the results are rather robust with respect to small variations of the solid angle.

\section{Test runs with no GW friction \label{eq:appA}}

\begin{figure}
    \centering
    \includegraphics[scale=0.6]{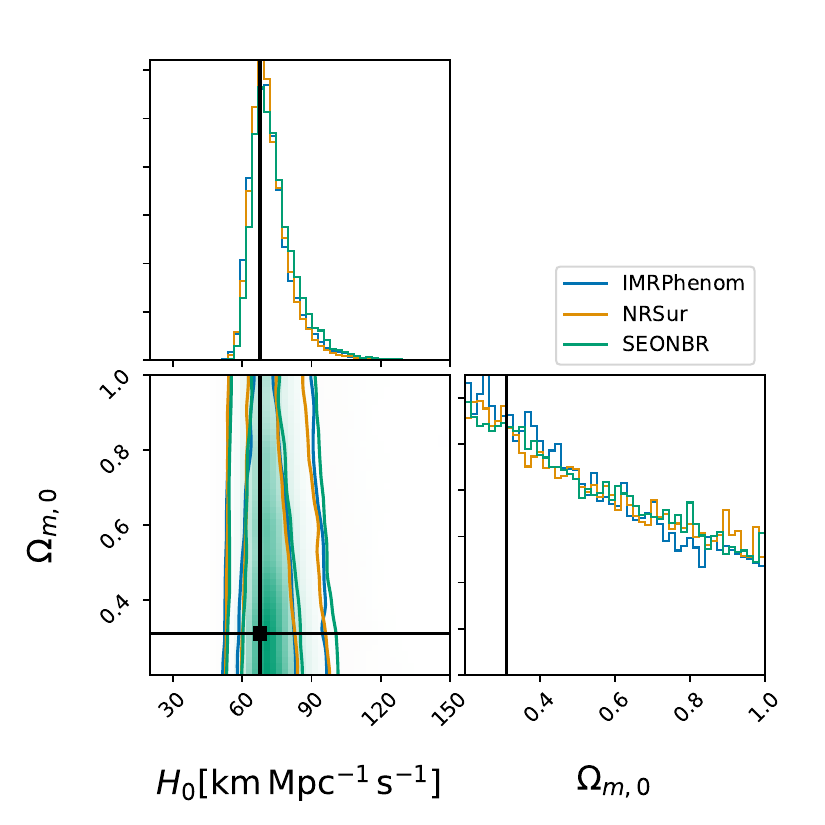}
    \caption{Posterior probability density function on $H_0$ and $\Omega_m$ for GW170817+GW190521, when assuming no GR deviation parameters. The posterior probability density function does not show any particular systematic when changing the waveform approximant as shown in \cite{Chen:2020gek}.}
    \label{fig:test_run}
\end{figure}

We ran an analysis fixing GR (no deviation parameters) and using priors on $H_0 \in [20,300] \hu $ and $\Omega_{m,0} \in [0.2,1.0]$ using the three different waveform models.
Fig.~\ref{fig:test_run} shows the posterior probability density function that we obtain on $H_0$ and $\Omega_{m,0}$.  
As it can be seen from the figure, the different waveform approximants give very similar estimation of the Hubble constant and $\Omega_{m,0}$ (lower values of $\Omega_{m,0}$ are preferred with some approximants).
We obtain a value of $H_0 = 74^{+13}_{-7} \hu$ and $\Omega_{m,0} =0.58^{+0.25}_{-0.25}$.
These results are consistent with the analysis of \cite{Mukherjee:2020kki} and \cite{Chen:2020gek}, where different priors were used to estimate $H_0$ and $\Omega_{m,0}$ jointly with the dark energy equation of state parameters.

\bibliographystyle{JHEP}
\bibliography{biblio}

\end{document}